\documentclass[conference]{IEEEtran}
\usepackage{cite}
\usepackage{amsmath,amssymb,amsfonts,amsthm}
\usepackage{algorithmic}
\usepackage{graphicx}
\usepackage{textcomp}
\usepackage{xcolor}
\usepackage{comment}
\usepackage{subfig}
\usepackage[ruled,vlined]{algorithm2e}
\usepackage{booktabs}
\usepackage{nicematrix}
\newtheorem{theorem}{Theorem}

\newtheorem{prop}{Proposition}
\newtheorem{remark}{Remark}
\newtheorem{defin}{Definition}

\def\BibTeX{{\rm B\kern-.05em{\sc i\kern-.025em b}\kern-.08em
    T\kern-.1667em\lower.7ex\hbox{E}\kern-.125emX}}
\begin{document}

\title{Energy-Efficient UAV Replacement in Software-Defined UAV Networks\\
}

 
\author{%
  \IEEEauthorblockN{Mohammad Javad-Kalbasi\IEEEauthorrefmark{1}
                    and Shahrokh Valaee\IEEEauthorrefmark{1}}
  \IEEEauthorblockA{\IEEEauthorrefmark%
                    {1}University of Toronto,
                    \{mohammad.javadkalbasi@mail, valaee@ece\}.utoronto.ca}

}
\IEEEoverridecommandlockouts
\maketitle
\IEEEpubidadjcol
\begin{abstract}
Unmanned Aerial Vehicles (UAVs) in networked environments face significant challenges due to energy constraints and limited battery life, which necessitate periodic replacements to maintain continuous operation. Efficiently managing the handover of data flows during these replacements is crucial to avoid disruptions in communication and to optimize energy consumption. This paper addresses the complex issue of energy-efficient UAV replacement in software-defined UAV network. We introduce a novel approach based on establishing a strict total ordering relation for UAVs and data flows, allowing us to formulate the problem as an integer linear program. By utilizing the Gurobi solver, we obtain optimal handover schedules for the tested problem instances. Additionally, we propose a heuristic algorithm that significantly reduces computational complexity while maintaining near-optimal performance. Through comprehensive simulations, we demonstrate that our heuristic offers practical and scalable solution, ensuring energy-efficient UAV replacement while minimizing network disruptions. Our results suggest that the proposed approach can enhance UAV battery life and improve overall network reliability in real-world applications.
\end{abstract}
\begin{IEEEkeywords}
Software-defined UAV network, energy-efficient UAV replacement, strict total ordering relation, integer linear program, Gurobi solver
\end{IEEEkeywords}


\section{\textbf{introduction}}
\begin{figure}[t]
\centerline{\includegraphics[height=5cm,width=9cm]{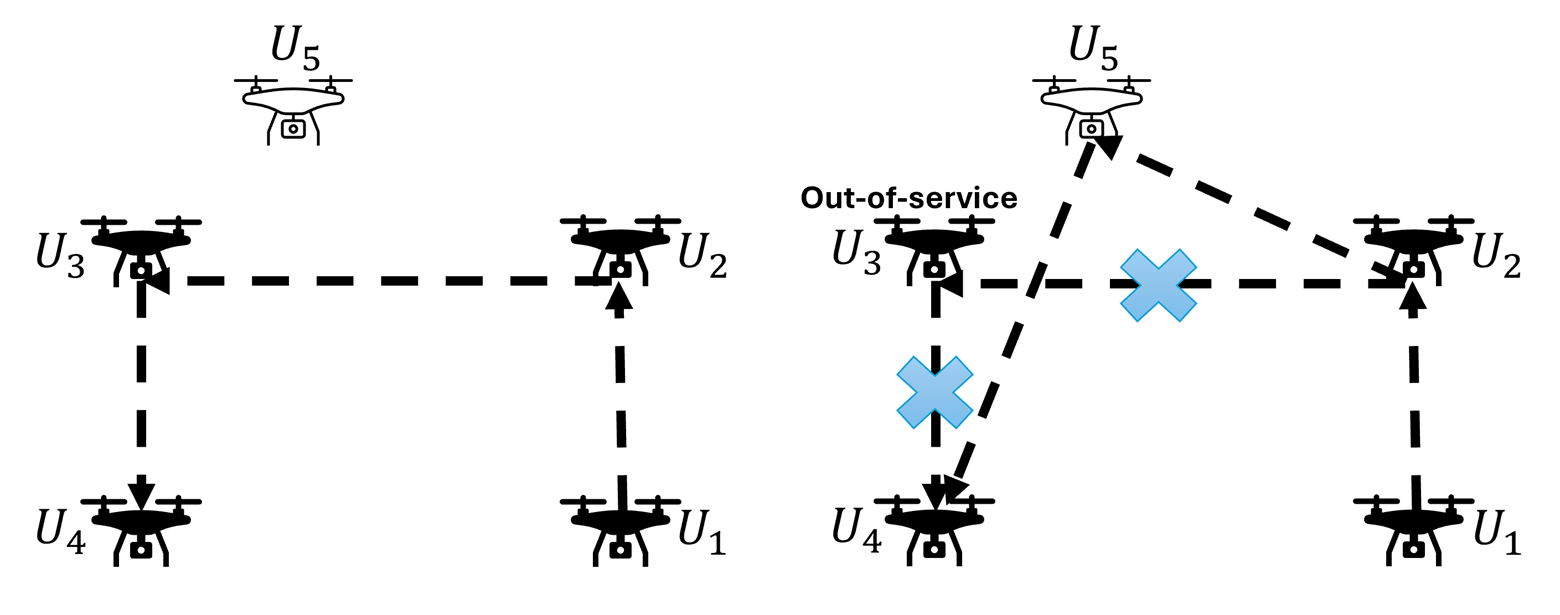}}
\caption{Illustration of a data flow handover in the UAV network.}
\label{typehand}
\end{figure}

Unmanned Aerial Vehicles (UAVs) demonstrate remarkable capabilities, including clear line-of-sight (LOS) ground node connectivity, rapid deployment, and adaptability, rendering them indispensable in a wide range of applications, including surveillance, search and rescue missions, delivery services, and communication networks. During a long mission, UAVs may periodically go out of service as they go out of power or develop faults \cite{pp1}. Their communication interfaces may also be shut down to conserve power, or one or more of the UAVs may be withdrawn, when less dense network is required. In all these cases the network needs to re-configure and the ongoing voice, video or data sessions are required to be handed over to one of the working UAVs according to some predefined criteria \cite{Lit00}. Handover allows for continuity of network communication with only a minor increase of message latency during the handover process \cite{pp2}. Fig. \ref{typehand} illustrates an example of a handover process. In this scenario, UAV \( U_3 \) runs out of power and must go out of service, requiring replacement by a new UAV, \( U_5 \), to extend the operational duration of network. Since the data flow \( U_1 \rightarrow U_2 \rightarrow U_3 \rightarrow U_4 \) currently passes through UAV \( U_3 \), it must be handed over to the new path \( U_1 \rightarrow U_2 \rightarrow U_5 \rightarrow U_4 \).

\subsection{Related work}
In the existing literature, researchers in \cite{lan2020development}-\cite{review2} have explored the concept of UAV replacement. Notably, \cite{lan2020development} and \cite{guetta2022onboard} have introduced a battery replacement system where UAV batteries are swapped at charging stations after landing, allowing them to resume communication services. However, this method introduces coverage gaps in the communication service area and falls short in ensuring uninterrupted communication for ground users. In contrast, \cite{sanchezaguero2020energyaware} proposed a scheduling algorithm to manage multiple UAV replacements, enabling long-term communication services, particularly during hotspot events, extending for several hours. Meanwhile, \cite{patra2019dynamic} introduced a model that involves swapping the positions of UAVs with depleted batteries with those of neighboring UAVs possessing higher remaining battery life. In \cite{bhola2021outagefree}, researchers investigated the issue of link outage during UAV replacements and formulated strategies to ensure outage-free replacements. In a different approach, \cite{shan2022multiuav} focused on the UAV charging mechanism by establishing multiple battery charging stations in the field, minimizing UAV energy consumption. However, that work did not prioritize meeting the communication service performance requirements of ground users or ensuring adequate data rates for each user. In \cite{review1} and \cite{review2}, the authors presented a framework that addresses the challenge of maintaining uninterrupted coverage in a UAV-assisted wireless communication system, particularly when the currently operating UAV depletes its energy reserves. In such scenarios, service continuity is ensured by substituting the exhausted UAV with a fully charged one. The primary goal of this replacement process is to maximize the total data rate attainable for all ground users. This objective is realized through the collaborative optimization of three-dimensional multi-UAV trajectories and resource allocation to individual users by the UAVs.



\subsection{Motivation}
The benefits of UAV networks come with significant challenges, especially when UAVs need to be taken out of service for reasons such as battery depletion, hardware malfunctions, or mission completion. In dense UAV networks, like the one depicted in Fig. \ref{statement}, UAVs are often responsible for maintaining important communication flows between multiple devices. When UAVs are scheduled to be replaced, the handover of communication flows passing through them must be carefully managed. Inefficient handovers can lead to increased energy consumption, as UAVs must hover for extended periods. This not only shortens the operational lifetime of the network but also negatively impacts mission objectives and service continuity.

The problem of minimizing the hovering energy consumption during the handover process is therefore crucial. Optimizing this process can extend the overall operational time of UAV networks, reduce operational costs, and improve the sustainability of these systems, especially in energy-constrained scenarios. Furthermore, energy-efficient UAV replacement is of particular importance in emergency response scenarios where prolonged aerial operation is essential for search-and-rescue missions or real-time data collection, and interruptions in service can have life-threatening consequences.

In addition to emergency response, other potential applications of energy-efficient UAV replacement include logistics and delivery systems, aerial surveillance for smart cities, and remote sensing operations in agriculture. These applications rely on the continuous and reliable operation of UAV networks over long periods, which makes the efficient replacement of UAVs a high-priority research problem. 
\subsection{Contribution}
In this paper, we examine how to replace UAVs efficiently in software-defined UAV networks while conserving energy. To the best of our knowledge, this problem has not been explored previously. To this end, we introduce a novel approach that centers around the establishment of a strict total ordering relation for UAVs and data flows. This approach enables us to formulate the problem as an integer linear program and leverage advanced tools such as the Gurobi solver to identify optimal handover schedules. Additionally, we propose a heuristic algorithm designed to significantly reduce computational complexity while maintaining efficiency.

\section{\textbf{Problem Statement}}\label{statsec}
In this paper, we analyze software-defined UAV networks that consist of a single SDN controller. The SDN paradigm is captivating as it injects flexibility and programmability into networks. SDN is a pivotal technology in the realization of 5G \cite{11sdn11}. It enables the operation and validation of networks in a more adaptable manner than traditional networks, allowing for the efficient implementation of customized services tailored to 5G networks \cite{33sdn33}. In the context of SDN, the controller is responsible for calculating configurations and transmitting forwarding rules to the corresponding switch. It is crucial that configuration updates are executed consistently and swiftly to prevent congestion, delays, and policy violations \cite{44sdn44}, \cite{pp3}.
A flow corresponds to a path created from a source node to a destination node in order to transmit traffic.  

\begin{figure}[t]
\centerline{\includegraphics[height=5cm,width=7.5cm]{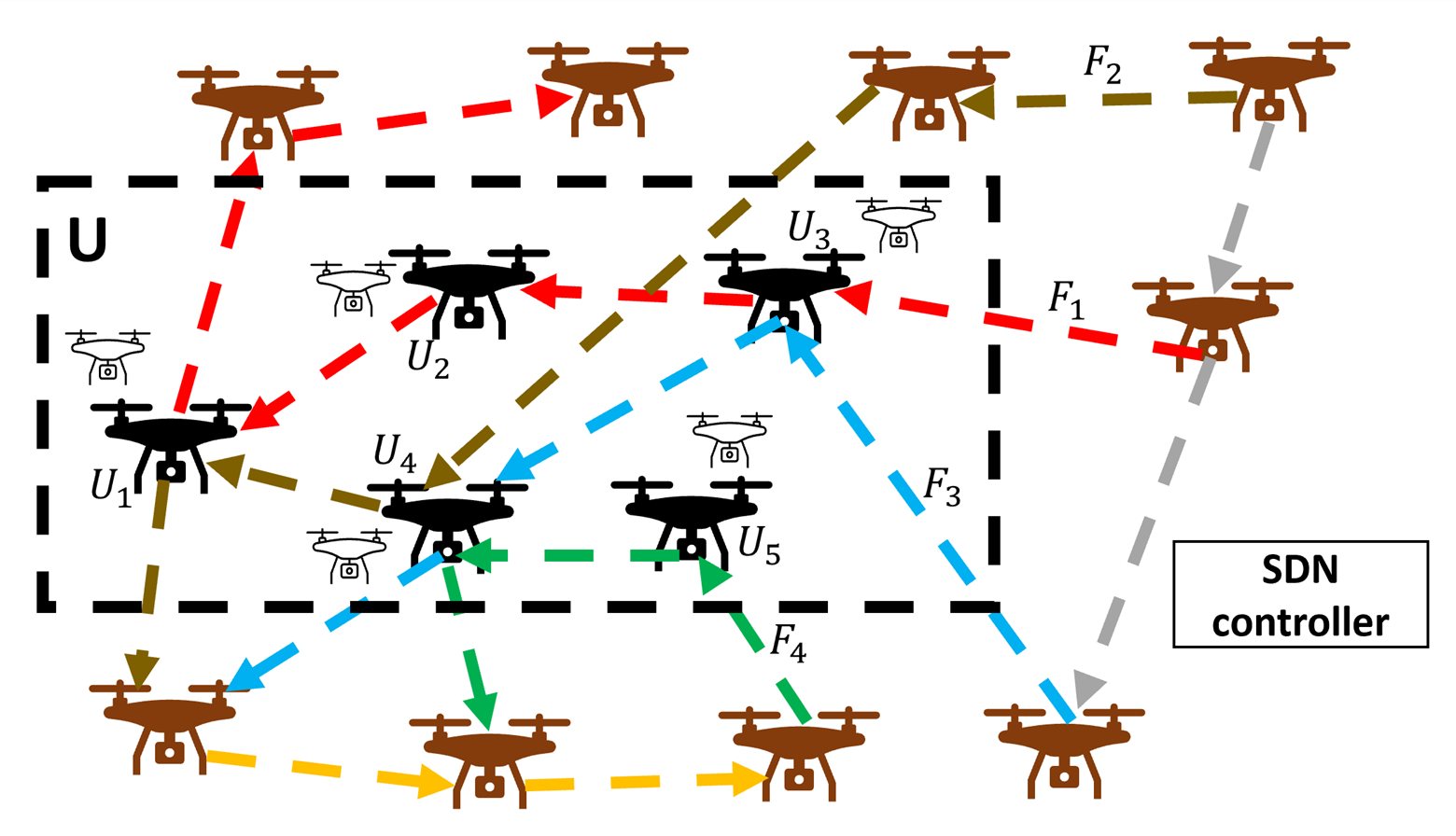}}
\caption{A software-defined UAV network.}
\label{statement}
\end{figure}

Let us consider a sample software-defined UAV network with $N_{u}$ UAVs and $N_{f}$ flows as given in Fig. \ref{statement}. We show the set of UAVs that need to go out of service with $U=\{U_{1},U_{2},\ldots,U_{m}\}$. Moreover, the set of flows that pass through at least one of the members of set $U$ is shown with $F=\{F_{1},F_{2},\ldots,F_{n}\}$. Although not all communication links between UAVs are shown in Fig. \ref{statement}, any flow from UAV \( U_{i} \) to UAV \( U_{j} \) indicates the presence of a communication link between them. Additionally, multiple flows can share the same link; however, this scenario is not depicted in Fig. \ref{statement} to enhance the visual clarity of different flow routes. We also point out that the new UAVs, taking the place of the old ones, are displayed in white. For the sample network in Fig.~\ref{statement}, we have $N_{u}=14$, $N_{f}=6$, $m=5$, and $n=4$. Let $R_{i}^{del}$, $R_{i}^{ins}$, and $R_{i}^{mod}$ be the number of deleted, inserted, and modified rules required for the handover of flow $F_{i}$. Moreover, we show the deletion, insertion, and modification time of each rule by $\tau^{del}$, $\tau^{ins}$, and $\tau^{mod}$, respectively. According to the test result in \cite{fcnrtime}, the delay of deleting and inserting is 5 ms, while the delay of modifying is set to 10 ms. Therefore, the required time for handover of flow $F_{i}$ (updating the control commands by the SDN controller) can be written as $T_{i}=~R_{i}^{del}\tau^{del}+~R_{i}^{ins}\tau^{ins}+R_{i}^{mod}\tau^{mod}$. For the handover scenario illustrated in Fig. \ref{typehand}, the forwarding rule \( U_2 \rightarrow U_3 \) must be modified to \( U_2 \rightarrow U_5 \), the rule \( U_3 \rightarrow U_4 \) needs to be deleted, and a new forwarding rule \( U_5 \rightarrow U_4 \) must be created. As a result, we have \( R_{i}^{del} = R_{i}^{ins} = R_{i}^{mod} = 1 \). Consequently, the total time required for the handover is given by \( T = \tau^{del} + \tau^{ins} + \tau^{mod} = 20 \) ms.

\begin{figure}[t]
    \centering
     \begin{minipage}[b]{0.8\columnwidth}
            \centering
            \includegraphics[width=1\linewidth, height=2.5cm]{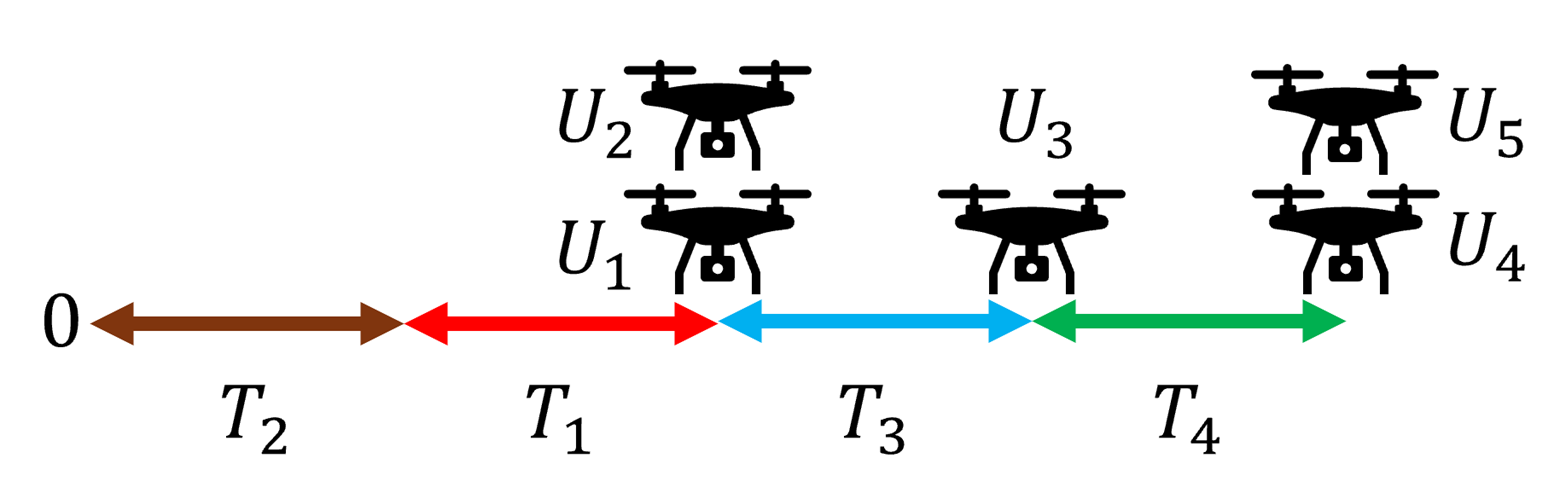} 
            \caption*{(a) schedule 1: $F_{2}$, $F_{1}$, $F_{3}$, and $F_{4}$}
    \end{minipage}
    \begin{minipage}[b]{0.8\columnwidth}
        \centering
        \includegraphics[width=1\linewidth, height=2.5cm]{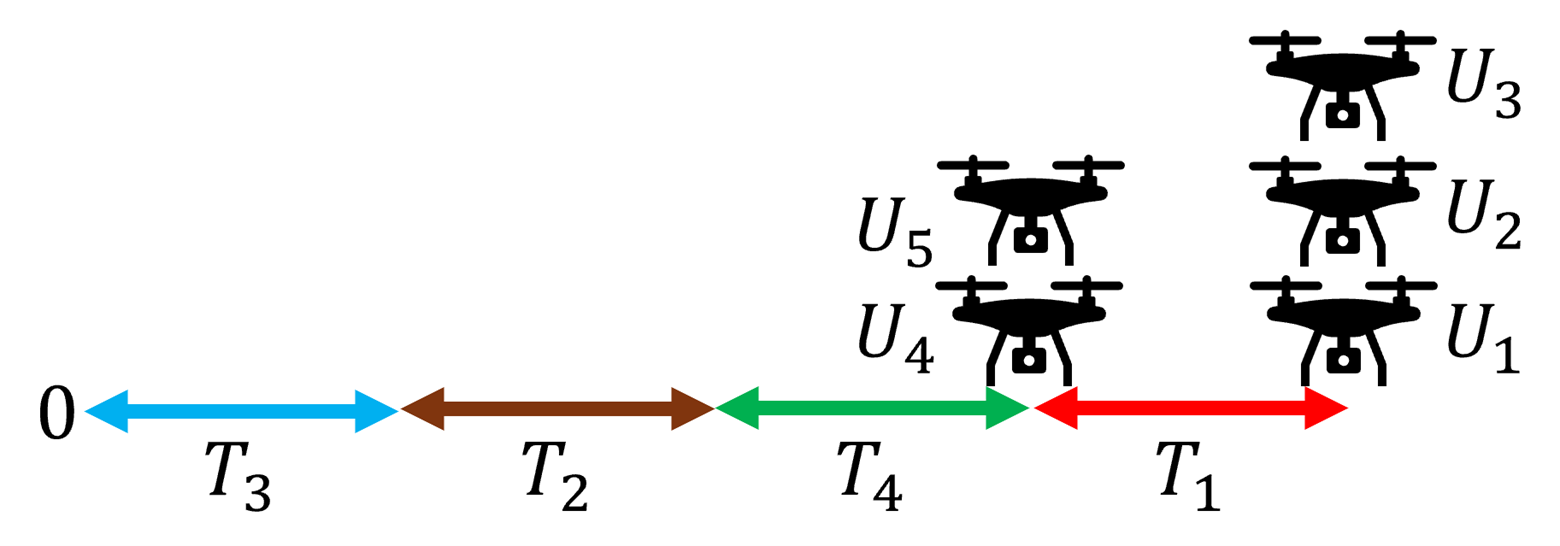} 
        \caption*{(b) schedule 2: $F_{3}$, $F_{2}$, $F_{4}$, and $F_{1}$}
    \end{minipage}
    \caption{Two handover schedules for the network in Fig. \ref{statement}.}
    \label{twoschedule}
\end{figure}

The primary focus of this study is to determine a handover schedule that minimizes the hovering energy consumption during the UAV replacement process. To see the effect of handover ordering on the energy efficiency, two different handover schedules for the sample network in Fig. \ref{statement} are schematically depicted in Fig. \ref{twoschedule}. The required time for handover of flow $F_{i}$ is shown with $T_{i}$ for $1\leq i \leq 4$. This figure shows the UAVs that go out of service at each time. We highlight that a UAV is considered out of service when all the flows passing through it are successfully handed over. For the handover ordering in schedule 1, UAVs $U_{1}$ and $U_{2}$ go out of service at time $T_{2}+T_{1}$. Next, UAV $U_{3}$ goes out of service at $T_{2}+T_{1}+T_{3}$. Finally, UAVs $U_{4}$ and $U_{5}$ stop operating at $T_{2}+T_{1}+T_{3}+T_{4}$. Hence, assuming that the hovering power of all UAVs is $P$, the hovering energy required for the first schedule is $E_{1}:=(5T_{1}+5T_{2}+3T_{3}+2T_{4})P$. In a similar way, we can show that the hovering energy consumption of the second schedule is $E_{2}:=(3T_{1}+5T_{2}+5T_{3}+5T_{4})P$. It is not hard to see that the handover times for the flows are $T_{1}=40$ ms, $T_{2}=30$ ms, $T_{3}=30$ ms, and $T_{4}=30$ ms. Therefore, for the special case of $P=100$ W, we have $E_{1}=50$ J and $E_{2}=57$ J, which shows that the first schedule achieves less hovering energy consumption compared to the second one. The central question now arises: how can one identify the optimal handover schedule with the least total hovering energy consumption?

\section{\textbf{Problem Formulation}}\label{formsec}

In this section, we introduce a systematic method that hinges on the establishment of a strict total ordering relation for UAVs and data flows. The objective is to address the problem of minimizing the total
hovering energy consumption during the UAV replacement in a software-defined UAV network. We begin by providing
some preliminary definitions that are necessary in formulating the problem \cite{relationbook}.
\begin{defin}
A relation $\mathbb{R}$ on the set $\mathbb{A}$ is defined as
\[\mathbb{R}=\{(a,b): a,b\in \mathbb{A}\}.\]
\end{defin}

The relation $\mathbb{R}$ can alternatively be shown by a directed graph in which $\mathbb{A}$ is the set of vertices and there is a directed edge from node $a$ to node $b$ if and only if $(a,b)\in \mathbb{R}$.
\begin{defin}
A relation $\mathbb{R}$ on the set $\mathbb{A}$ is called a strict total ordering relation if the following properties hold
\begin{itemize}
    \item $\mathbb{R}$ is irreflexive: $(a,a)\notin \mathbb{R}$ for every $a \in \mathbb{A}$.
    \item Any two members of $\mathbb{A}$ are comparable: $(a,b)\in \mathbb{R}$ or $(b,a)\in \mathbb{R}$.
    \item $\mathbb{R}$ is asymmetric: $(a,b)\in \mathbb{R}$ implies that $(b,a)\notin \mathbb{R}$.
    \item $\mathbb{R}$ is transitive: $(a,b)\in \mathbb{R}$ and $(b,c)\in \mathbb{R}$ imply that $(a,c)\in \mathbb{R}$. 
\end{itemize}
\end{defin}

A strict total ordering relation imposes a strict hierarchy on the elements of a set, ensuring that no two distinct elements are considered equal. It is not hard to see that the corresponding graph of a strict total ordering relation is a complete directed graph in which there is only one directed edge between any two nodes in the graph. Moreover, the longest path in the graph induces a valid ordering for the set of nodes. 

To simplify the notation in our formulation, we map the combined set \( F \cup U \) onto a new set, denoted as \( \mathcal{K} =~\{K_1, K_2, \ldots, K_{n+m}\} \). Specifically, each \( F_i \) is assigned to \( K_i \) for \( 1 \leq i \leq n \), and each \( U_j \) is assigned to \( K_{n+j} \) for \( 1 \leq j \leq m \). This mapping is introduced because we aim to establish a strict total ordering for both UAVs and data flows. The combined set \( \mathcal{K} \) is thus essential and serves as a foundational component in the problem formulation.

\begin{defin}
For a given handover problem instance (such as problem instance in Fig. \ref{statement}), we define the dependency relation $\mathbb{D}$ on the set $\mathcal{K}$ in which $(K_{i},K_{j})\in \mathbb{D}$ if and only if flow $F_{i}\in F$ passes through UAV $U_{j-n} \in U$. 
\end{defin}

\begin{figure}[t]
\centerline{\includegraphics[height=3cm,width=7cm]{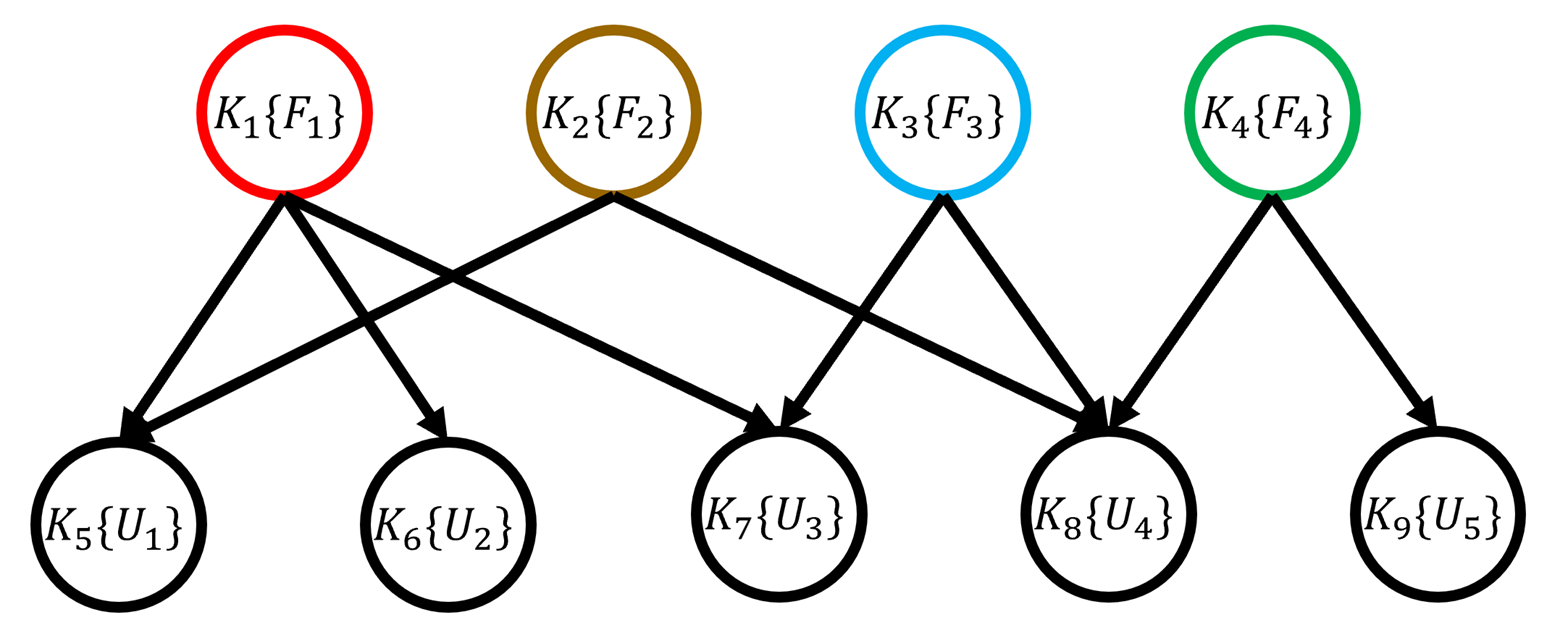}}
\caption{Dependency graph of network given in Fig. \ref{statement}.}
\label{Dependencii}
\end{figure}
Dependency relation $\mathbb{D}$ demonstrates the fact that a UAV can go out of service as soon as all the flows passing through it have
been handed over. Hence, the corresponding graph of a dependency relation is always bipartite. For an instance, the dependency graph of sample problem
in Fig. \ref{statement} is depicted in Fig. \ref{Dependencii}. As it can be seen, there are directed edges from $K_{1}\{F_{1}\}$ to $K_{5}\{U_{1}\}$, $K_{6}\{U_{2}\}$, and $K_{7}\{U_{3}\}$ because the first flow ($F_{1}$) passes through the UAVs $U_{1}$, $U_{2}$, and $U_{3}$.

In what follows, we formally present our approach to minimize the hovering energy consumption of UAVs that go out of service during the replacement process. Our method is based on finding a strict total ordering relation $\mathbb{R}$ on the set $\mathcal{K}$ for which $\mathbb{D}\subseteq \mathbb{R}$. We point out that such relation induces an ordering for the set $F$ as well, which is indeed a handover schedule. Let $[n+m]:=\{1,2,\ldots,n+m\}$. 
\begin{figure}[t]
\centerline{\includegraphics[height=5cm,width=5cm]{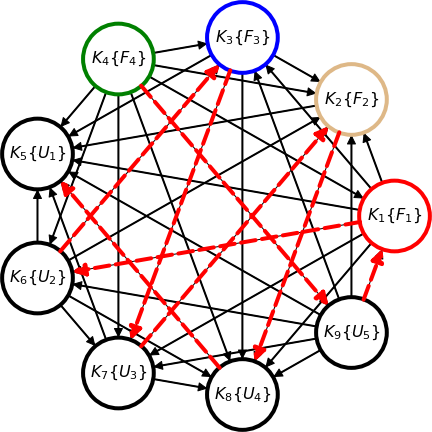}}
\caption{Optimal strict total ordering relation for the sample problem given in Fig. \ref{statement} and induced hierarchy $F_{4}\rightarrow~U_{5} \rightarrow~F_{1} \rightarrow U_{2} \rightarrow F_{3} \rightarrow U_{3} \rightarrow F_{2} \rightarrow U_{4} \rightarrow U_{1}$ shown in red dashed arrows.}
\label{optimalordering}
\end{figure}
To simplify the notation, we implicitly assume hereafter that tuples and sets have no multiplicity. Therefore, $(a_{1},\ldots,a_{l})\in {[n+m]}^{l}$ and $\{b_{1},\ldots,b_{l}\}\subseteq [n+m]$ denote a tuple and a set, respectively, with $l$ distinct elements. For formulating the problem, we define the binary indicator variable $x_{ij}$ that takes the value $1$ if $(K_{i},K_{j})\in \mathbb{R}$ and $0$, otherwise. Since $\mathbb{D}\subseteq \mathbb{R}$, we conclude that $x_{ij}=1$ for $(K_{i},K_{j})\in \mathbb{D}$. Moreover, from the facts that any two members of $\mathcal{K}$ are comparable and $\mathbb{R}$ is asymmetric, we have $x_{ij}+x_{ji}=1$ for each $\{i,j\}\subseteq [n+m]$. Additionally, transitivity of $\mathbb{R}$ implies that $x_{ij}+x_{jk}\leq 1+x_{ik}$ for each $(i,j,k) \in {[n+m]}^{3}$. Finally, assuming that the hovering power of UAV $U_{l}$ is $P_{l}$ for $1 \leq l \leq m$, the total hovering energy consumption of UAVs that go out of service during the replacement process can be written as $\sum\limits_{i=1}^{n}\sum\limits_{j=n+1}^{n+m}T_{i}x_{ij}P_{j-n}$. Therefore, our goal problem can be formulated as the following integer linear program
\begin{equation}\label{goaloptima}
\left\{ \begin{array}{ll}
\min\limits_{x_{ij}}\sum\limits_{i=1}^{n}\sum\limits_{j=n+1}^{n+m}T_{i}x_{ij}P_{j-n}\\\\
x_{ij}\in \{0,1\}~~\text{for}~~(i,j)\in {[n+m]}^{2}\\\\
x_{ij}=1 ~~\text{for}~~(K_{i},K_{j})\in \mathbb{D} \\\\
x_{ij}+x_{ji}=1~~\text{for}~~\{i,j\}\subseteq [n+m]\\\\
x_{ij}+x_{jk}\leq 1+x_{ik}~~\text{for}~~(i,j,k)\in {[n+m]}^{3}
\end{array}
\right.
\end{equation}
As it can be seen, the complexity of the optimization problem (\ref{goaloptima}) escalates notably as both $n$ and $m$ increase, resulting in a substantial increase in the number of binary variables and constraints. We emphasize that every viable solution to optimization problem (\ref{goaloptima}) constitutes a strict total ordering relation on the combined set $F \cup U$. Furthermore, the optimal solution to this problem establishes a hierarchy with the least total hovering energy consumption. As an illustration, Fig. \ref{optimalordering} displays the optimal solution and induced hierarchy for the sample problem presented in Fig. \ref{statement}. This solution is obtained by solving the optimization problem (\ref{goaloptima}) with the Gurobi solver. As evident, the corresponding handover schedule is $F_{4}$, $F_{1}$, $F_{3}$, and $F_{2}$. Consequently, the optimal hovering energy consumption is given by $E^{\text{opt}} := (4T_{1} +~2T_{2} +~3T_{3} +~5T_{4})P$. Substituting the values provided in Section \ref{statsec}, we find $E^{\text{opt}} =~46$ J, which is lower than the hovering energy consumption of the schedules presented in Fig. \ref{twoschedule}.


\begin{algorithm}[t]
\caption{Proposed heuristic algorithm}
\begin{algorithmic}
\REQUIRE $T_{i}$, $\Delta_{i}$, $P_{j}$, and $\Lambda_{j}$.
\STATE \textbf{1)} Compute $H_{j}$ for $1 \leq j \leq m$.
\STATE \textbf{2)} Compute $S_{i}$ for $1 \leq i \leq n$.
\STATE \textbf{3)} Sort $S_{i}$ values in descending order.
\STATE \textbf{4)} Hand over the flows based on the obtained order.
\end{algorithmic}
\end{algorithm}
\section{\textbf{Proposed Heuristic Algorithm}}\label{heusec}
In this section, we present a heuristic algorithm designed for the optimization problem (\ref{goaloptima}). 
For \( 1 \leq i \leq n \), let \( \Delta_{i} \) represent the set of UAVs in \( U \) that are traversed by flow \( F_{i} \). Additionally, we define \( \Lambda_{j} \) as the set of flows passing through the \( j \)-th UAV, where \( 1 \leq j \leq m \). 

Furthermore, we introduce \( H_{j} \triangleq \sum\limits_{i \in \Lambda_{j}}T_{i} \), which denotes the total processing time of all flows within \( \Lambda_{j} \). Essentially, \( H_{j} \) represents the time required for the replacement of the \( j \)-th UAV.

With these definitions established, we can now define a score for each flow \( F_{i} \), which will be instrumental in formulating our proposed heuristic algorithm.

\begin{defin}\label{score}
The score of the $i$-th flow $F_{i}$ is defined as
\[S_{i}\triangleq \sum\limits_{j \in \Delta_{i}}\frac{P_{j}}{H_{j}}.\]
\end{defin}

We point out that the unit of defined score is $\frac{J}{s^{2}}$. Noting Definition \ref{score}, one can expect that a flow with high score is passing through the UAVs with high hovering powers which require less time for the replacement. As a result, those flows with high scores are good candidates to be handed over first to save the hovering energy consumption of UAVs. The formal description of proposed heuristic algorithm is given in Algorithm 1. In Algorithm 1, we initially calculate score values for all flows, and subsequently, these values are arranged in descending order to establish the handover sequence. 

\begin{prop}\label{lemcomplex}
Algorithm 1 has a time complexity of $\mathcal{O}\left(n\log(n)+\sum\limits_{i=1}^{n}|\Delta_{i}|+\sum\limits_{j=1}^{m}|\Lambda_{j}|\right)$.
\end{prop}
\proof We determine the overall time complexity by evaluating the time complexity of each individual step. Considering the structure of Algorithm 1, we can conclude that calculating the values of $H_{j}$ for $1 \leq j \leq m$ in the first step has a time complexity of $\mathcal{O}\left(\sum\limits_{j=1}^{m}|\Lambda_{j}|\right)$. In the second step, computing the values of $S_{i}$ for $1 \leq i \leq n$ has a time complexity of $\mathcal{O}\left(\sum\limits_{i=1}^{n}|\Delta_{i}|\right)$. Finally, sorting the $S_{i}$ values in the third step has a time complexity of $\mathcal{O}(n\log(n))$. Therefore, the overall time complexity of Algorithm 1 is $\mathcal{O}\left(n\log(n) + \sum\limits_{i=1}^{n}|\Delta_{i}| + \sum\limits_{j=1}^{m}|\Lambda_{j}|\right)$.\qed

As an illustration, we addressed the sample problem presented in Fig. \ref{statement} using Algorithm 1. Consequently, the resulting handover schedule is $F_{1}$, $F_{4}$, $F_{3}$, and $F_{2}$. It is evident that the corresponding hovering energy consumption for this schedule is $E^{\text{Algorithm 1}} :=~(5T_{1} +~2T_{2} +~3T_{3} + 4T_{4})P$. Upon substituting the values provided in Section \ref{statsec}, we find $E^{\text{Algorithm 1}} = 47$ J, only $1$ J more than $E^{\text{opt}}$.

\section{\textbf{Numerical Results}}\label{numsec}
This section demonstrates the numerical results. To implement the suggested heuristic algorithms, Python 3.8.5 was employed. Additionally, for the optimization problem (\ref{goaloptima}), we utilized the Gurobi solver to obtain the optimal solution.

\subsection{\textbf{Simulation Setup}}
For the simulation, we considered a software-defined UAV network with one SDN controller and $N_{u}=40$ UAVs which were randomly distributed in a square area of $150$ m $\times$ $150$~m. Since UAVs hover at high altitude, they maintain LOS channel between each other \cite{8613833}. Hence, the path loss between the $u$-th and the $v$-th UAVs can be expressed as $\Gamma_{u,v}=20\log\left(\frac{4 \pi f_{c}d_{u,v}}{c}\right)$, where $f_{c}$ is the carrier frequency, $c$ is the light speed, and $d_{u,v}$ is the distance between the $u$-th and the $v$-th UAVs \cite{saifglobecom}. Therefore, the SNR in dB between the $u$-th and the $v$-th UAVs can be written as $\gamma_{u,v}=10\log\left(p\right)-\Gamma_{u,v}-10\log\left(N_{0}\right)$, where $p$ is the transmit power of the $u$-th UAV, which is maintained fixed for all the UAVs, and $N_{0}$ is the additive white Gaussian noise variance. We assume that the $u$-th UAV and the $v$-th UAV have a successful link provided that $\gamma_{u,v}\geq \gamma_{0}$, where $\gamma_{0}$ is the minimum SNR threshold for the communication link between the UAVs \cite{saifglobecom_E}. The hovering power of the $j$-th UAV can be expressed by $P_{j}=\sqrt{\frac{{{(M_{j}g)}}^{3}}{2\pi r_{p}^{2}n_{p}\rho}}$, where $M_{j}$ is the mass of the $j$'th UAV, $g$ is the gravitational acceleration of the earth, $r_{p}$ is propellers’ radius, $np$ is the number of propellers, and $\rho$ is the air density \cite{javadpaperi}. In this simulation, we assume that $M_{j}$ is randomly chosen from the set $\{1, 2, 3, 4, 5\}$ kg. The total number of flows ($N_{f}$) was set to $70$ and $100$. Furthermore, we made the assumption that a portion ranging from 12.5\% to 25\% (specifically, when $m$ takes on values in the set $\{5, 6, 7, 8, 9, 10\}$) of the UAVs must be taken out of service and substituted with new ones. The values of the simulation parameters not explicitly mentioned here are the same as those used in \cite{saifglobecom_E} and \cite{javadpaperi}. 
\begin{figure}[t]
\centerline{\includegraphics[width=1\linewidth]{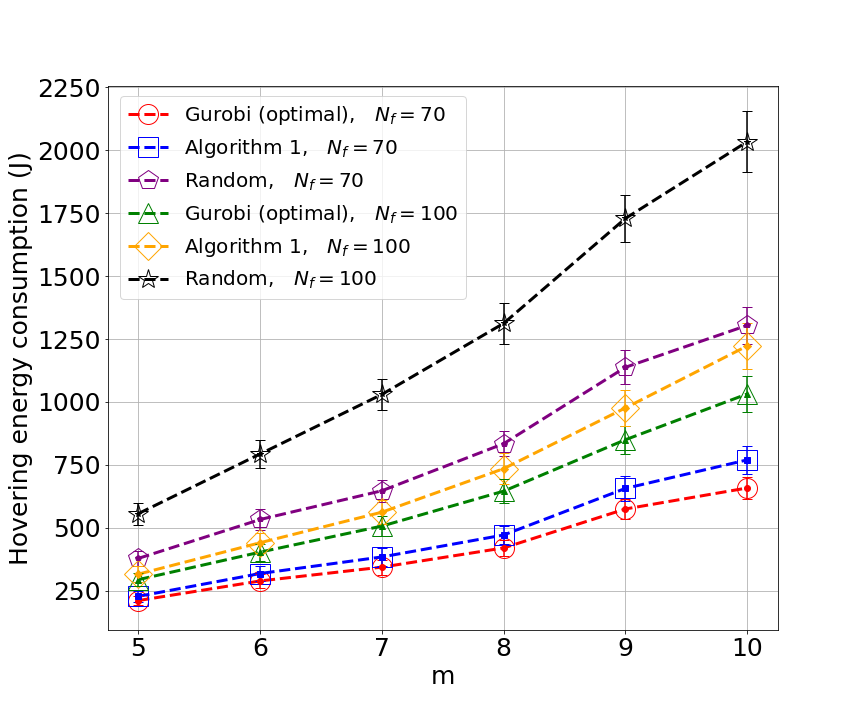}}
\caption{Hovering energy consumption versus $m$.}
\label{hovering energy}
\end{figure}
\subsection{\textbf{Performance Evaluation}}
To generate data flows, we randomly select two distinct UAVs and establish a route based on the shortest path between them. Since deriving a closed-form expression for the true expected value of hovering energy consumption is infeasible, we employ Markov Chain Monte Carlo (MCMC) simulations to approximate this expectation. In each iteration of the MCMC simulation, a random set of data flows is generated within the network, and the hovering energy consumption \(E_k\) is computed for all \( k = 1,2,\ldots,K \). The MCMC process is repeated for a sufficiently large number of iterations to ensure a comprehensive exploration of the parameter space. The sample mean of hovering energy consumption is given by $\overline{E} \triangleq \frac{1}{K} \sum\limits_{k=1}^{K} E_{k}$, which serves as an approximation of the true expected value. To assess the accuracy of this approximation, we compute an approximate 95\% confidence interval for the true expected value using the standard error (SE) of the MCMC samples, defined as  $\text{SE} = \frac{1}{\sqrt{K}} \sqrt{\frac{1}{K-1} \sum\limits_{k=1}^{K} \left(E_{k} - \overline{E}\right)^{2}}$ \cite{montebook}. The corresponding 95\% confidence interval is then approximated as  
$\text{CI} = \left(\overline{E} - 1.96 \times \text{SE}, \overline{E} + 1.96 \times \text{SE} \right)$ \cite{montebook}. This confidence interval estimates the range within which the true expected value of hovering energy consumption is likely to fall provided that the sample size \(K\) is sufficiently large.

Fig. \ref{hovering energy} presents data on the average energy consumption for different values of \(N_{f}\). The confidence intervals are depicted as vertical error bars around each data point. The results are quite evident: the proposed methods yield substantial improvements when compared to a random handover schedule. Specifically, for the case when $m=10$, the proposed methods achieve approximately half the energy consumption of the random schedule, marking a significant improvement. It is worth noting that Algorithm~1 closely approaches the optimal performance, underscoring its effectiveness in handling the handover schedule. Fig. \ref{hovering energy} also illustrates how $N_{f}$ and $m$ impact the level of hovering energy consumption. As expected, the energy consumption rises as $N_{f}$ and $m$ increase because more flows need to be handed over and more UAVs need to be replaced during the replacement process. We emphasize that in long missions requiring multiple phases of UAV replacements, the reduction in hovering energy consumption becomes even more significant compared to a random scheduling approach.

\begin{figure}[t]
\centerline{\includegraphics[width=1\linewidth]{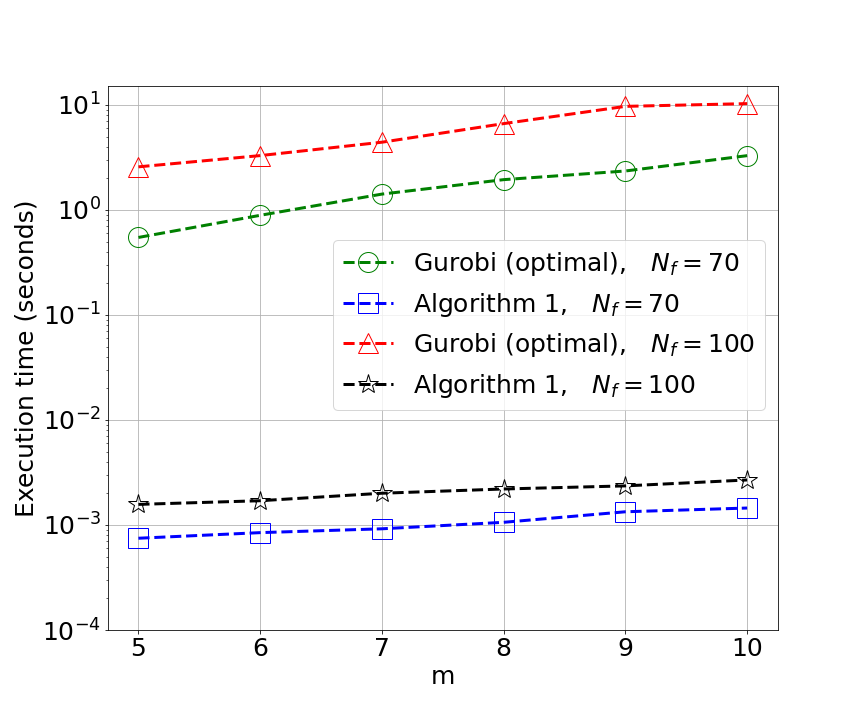}}
\caption{Execution time versus $m$}
\label{executiontime}
\end{figure}

We also conducted a comparison of execution times for various methods, as depicted in Fig. \ref{executiontime}. As evident from the data, Algorithm 1 exhibits execution time on the order of milliseconds, whereas the Gurobi solver, employed to find the optimal handover schedule, operates on the order of seconds. Furthermore, as anticipated, the execution time of all the proposed methods increases with the growth of $N_{f}$ because more flows need to be handed over.

\section{\textbf{Conclusion}}\label{consec}
UAVs encounter a significant challenge arising from energy constraints, often leading to the need for replacements. Effectively managing the handover of data flows during these replacements is crucial to avoiding interruptions in information transmission and minimizing energy consumption. This paper introduced an innovative approach to address the problem of energy-efficient UAV replacement in software-defined UAV networks by formulating it as an integer linear program. Our proposed method centers around establishing a strict total ordering relation for both UAVs and data flows.

Furthermore, we presented an efficient heuristic algorithm designed to mitigate the time complexity associated with solving the problem. Simulation experiments demonstrated that this heuristic closely approximates optimal solution while significantly reducing the computational burden of solving the problem.

\end{document}